\pdfoutput=1
\documentclass[aps,preprint,showkeys,amsmath,amssymb]{revtex4-2}
\usepackage{graphicx}
\begin{document}
\raggedbottom

\title{Dispersion engineering of infrared epsilon-near-zero modes by strong coupling to optical cavities}

\author{Ben Johns}
\email[]{benjohns@iisermohali.ac.in}
\affiliation{%
	Department of Chemical Sciences, Indian Institute of Science Education and Research, Mohali, India, 140306
}%


\date{30 March, 2023}
	
\begin{abstract} 
Epsilon-near-zero (ENZ) materials have recently emerged as a promising platform for infrared nanophotonics. A significant challenge in the design of ENZ-based optics is to control the dispersion of ENZ modes,  which otherwise have a flat profile near the ENZ frequency.
Strong coupling with an optical cavity is a promising approach to ENZ dispersion engineering, which however has limitations due to the lack of tunability or nanofabrication demands of the cavity employed.
Here, we theoretically and numerically show that much of the limitations of previous demonstrations can be overcome by strongly coupling the ENZ mode to an unpatterned Fabry-Perot cavity.
We demonstrate this unprecedented ENZ dispersion control in coupled cavities by designing tunable infrared polarizers that can absorb \textit{s} and reflect \textit{p}-polarized components, or vice versa, for almost any oblique angle of incidence, i.e. omnidirectional polarizers. The feasibility of active control is also demonstrated using a phase change material within the cavity, which predicts dynamic switchability of polariton dispersions across multiple resonant levels at mid-infrared wavelengths. These results are expected to advance the current understanding of strongly coupled ENZ interactions and demonstrate their potential in tailoring dispersions for active and passive control of light.
\end{abstract}

\maketitle	
	
\section{Introduction} 

Coherent optical interactions lead to unusual phenomena such as Fano resonances \cite{limonov2017fano}, electromagnetically induced transparency \cite{peng2014and}, extraordinary optical transmission \cite{rodrigo2016extraordinary}, surface lattice resonances \cite{bin2021ultra}, and strong coupling \cite{garcia2021manipulating}. Strong coupling (SC) is characterized by a splitting of resonances when the frequencies of mutually coupled resonators or cavities are brought close to each other. This effect is typified by an anti-crossing of the cavity dispersions, which split to form two distinct (upper and lower) polariton branches \cite{novotny2010strong}. The frequency separation at the anti-crossing point, known as Rabi splitting, signifies the strength of the mutual interaction, and SC can be experimentally observed in optical systems where the Rabi splitting is larger than the line-widths of individual resonances \cite{thomas2018plexcitons,dovzhenko2018light,hummer2013weak,schlather2013near,bhatt2021enhanced,lather2022cavity}. As a result of this splitting, SC opens up unique possibilities in engineering spectral response and dispersion in nanophotonics by finely tailoring polariton dispersions.

The introduction of SC has been proposed as an effective tool for exploiting the unique properties of epsilon-near-zero (ENZ) materials \cite{bruno2020negative, campione2015epsilon}. ENZ materials have a vanishing or near-zero permittivity ($\epsilon \to$ 0) at a particular wavelength and have attracted interest as a novel platform for exotic light-matter interactions \cite{niu2018epsilon}. Near the zero-epsilon wavelength ($\lambda_{ZE}$), ENZ materials can exhibit extreme field concentration and enhancement, strong optical non-linearities and perfect absorption \cite{silveirinha2007theory,alam2016large,jin2011arbitrarily,enoch2002metamaterial,xu2021broadband,johns2020epsilon}. Moreover, the optical modes supported in ultrathin ENZ films, called ENZ modes \cite{vassant2012berreman}, show strong field enhancements within low mode volumes, and have been utilized in active opto-electronic devices and ultrafast optical modulation \cite{vassant2012epsilon,feng2012coherent,tyborski2015ultrafast,yang2019high,johns2022tailoring}. 
Notably, the flat spectral dispersion near $\lambda_{ZE}$ results in a zero group velocity, low propagation lengths, and it overall limits control over the operating frequencies or angles \cite{runnerstrom2018polaritonic}. Earlier efforts in spectral shaping and tailoring of the ENZ mode dispersion explored its strong coupling with optical cavities such as quantum wells \cite{campione2015epsilon}, gap plasmon modes \cite{hendrickson2018experimental}, metasurfaces \cite{jun2013epsilon}, phonon polaritons in polar dielectrics \cite{passler2018strong}, plasmonic nanoantennas \cite{schulz2016optical,habib2020controlling,ghindani2021gate} and plasmon polaritons \cite{runnerstrom2018polaritonic}, which have helped realize negative refraction \cite{bruno2020negative}, hybrid plasmonic modes \cite{runnerstrom2018polaritonic}, and enhanced optical non-linearities \cite{alam2018large}. However, the ability to tailor dispersions in strongly coupled systems is often hampered by the low tunability of the employed optical cavities. Moreover, the use of metasurfaces and plasmonic resonances involve nanofabrication processes that increase the cost and complexity of the system. Therefore, planar, lithography-free structures with large field enhancements and strong tunability need to be further explored as potential platforms to exploit strong coupling with ENZ modes.

Here, we overcome the limitations of previous strongly coupled ENZ designs to demonstrate unprecedented tailoring of polariton dispersions at infrared wavelengths by coupling to a Fabry-Perot (FP) cavity. FP cavities have long been the workhorse of strong coupling research, having been employed in polariton-enhanced transport \cite{orgiu2015conductivity}, chemical reactivity \cite{thomas2019tilting}, and condensation \cite{plumhof2014room}; however, to the best of our knowledge, coupling of ENZ modes to a FP cavity has not been addressed earlier. Using an analytical approach, we identify the factors that control the coupling strength of the two modes, revealing that the polariton properties are mediated by an interplay of the near field of the ENZ and FP cavities as well as the ENZ thickness. In particular, we demonstrate the potential for ENZ dispersion engineering and complex spectral shaping in our system by designing planar, multilayer, coupled cavities that act as angle-independent, nearly omnidirectional polarizers with a controllable operation wavelength in the near and mid-infrared. Finally, to demonstrate their potential applications, active tunability of multilevel polariton dispersions is explored employing a phase change material. Our results not only demonstrate and characterize the extent to which ENZ mode dispersions can be engineered but also provide a simple configuration where ENZ light-matter interactions can be readily tailored for the versatile design of infrared optical components \cite{shahsafi2020infrared} and thermal emission control \cite{ hwang2022simultaneous}.

\section{\label{RnD}Results and discussion}
\subsection{Strong coupling of ENZ and FP modes}

\begin{figure*}
	\includegraphics[width=\textwidth]{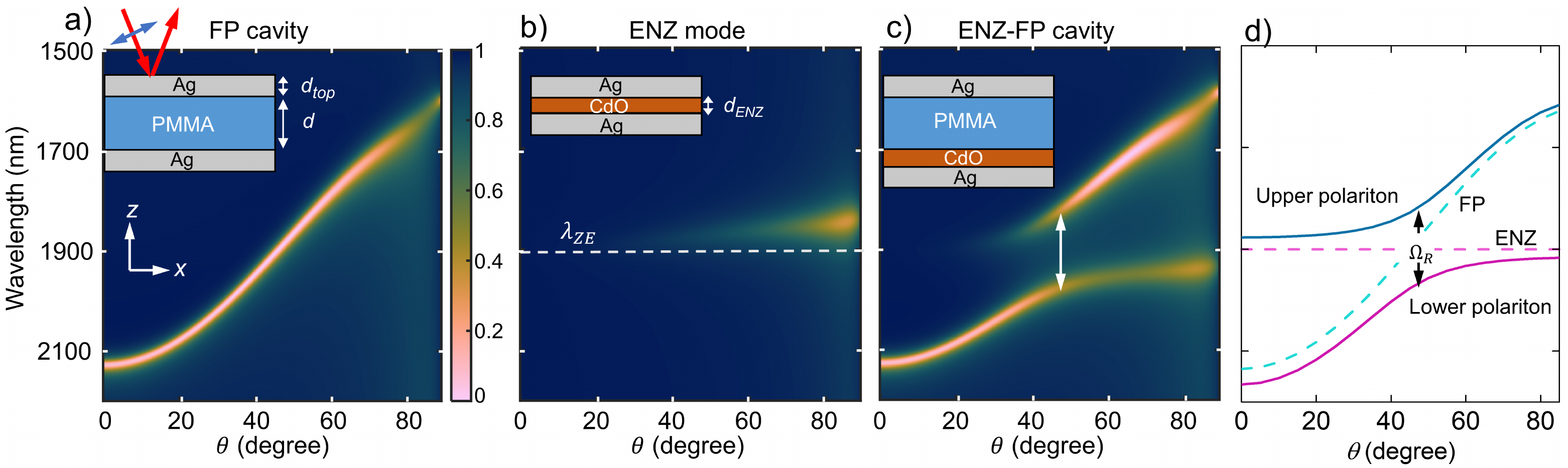}%
	\caption{(a-c) Simulated reflectance color maps for \textit{p}-polarized light incident on (a) FP cavity, (b) ENZ mode structure, and (c) ENZ-FP cavity. The corresponding geometries and coordinate axes are shown schematically as inset. The bottom Ag layer is semi-infinite while the top-most medium is assumed to be air in each case. The color scale is common in (a-c). (d) Calculated dispersion relations of the individual FP and ENZ modes (dashed curves) and the upper and lower branches of the strongly coupled dispersion (solid curves). The y-axis is common for (a-d). Arrows in (c,d) indicate the extent of splitting at the crossing point. \label{Fig:f1}}
\end{figure*}

To investigate the possibility of strongly coupled resonances in a planar geometry, a structure composed of a dielectric (PMMA) and an ENZ layer (doped cadmium oxide, CdO) sandwiched between two metallic mirrors (Ag) is considered. The permittivity or refractive index data used in this work and their model parameters are given in section 'Materials and Methods' and plotted in Figure S1 in Supplementary Materials. The individual or 'bare' resonances i.e. with only the dielectric between the mirrors (FP cavity: Ag-PMMA-Ag) and only the ENZ medium between the mirrors (ENZ mode: Ag-CdO-Ag) are first separately analyzed. The inset to Figure \ref{Fig:f1}a schematically shows the FP structure, where the PMMA dielectric layer (thickness $d$) is sandwiched between the Ag substrate and a thin Ag layer of thickness $d_{top}$. The thickness of the top metallic layer is comparable to its skin depth to allow coupling of light incident from the top-most air medium into the FP cavity. Throughout this work, the Ag substrate  is assumed to be semi-infinite with transmittance $T$ = 0. 
Figure \ref{Fig:f1}a shows the calculated color map of reflectance ($R$) of the FP cavity for \textit{p}-polarized light, i.e. $R_p$, as a function of angle of incidence ($\theta$) and wavelength ($\lambda$) in the near-IR (1500 nm to 2200 nm). The thicknesses are set as $d$ = 670 nm and $d_{top}$ = 20 nm. The dispersion of the FP cavity is evident as a sharp dip in reflectance (bright region), and the wavelength of maximum absorption ($A$ = $1-R$) varies over approximately a 500 nm range as $\theta$ is varied from 0 to 90 degrees in Figure \ref{Fig:f1}a. 
Figure \ref{Fig:f1}b shows $R_p$ for light incident from air onto the ENZ mode structure. Here, a CdO layer with a $\lambda_{ZE}$ of 1900 nm and thickness $d_{ENZ}$ = 20 nm is sandwiched between the Ag mirrors (see inset). The plot shows a flat dispersion in the vicinity of $\lambda_{ZE}$, indicating the excitation of the radiative ENZ mode known as Berreman mode \cite{vassant2012berreman}. 
Importantly, the dispersions of the FP cavity and the ENZ mode can cross near $\lambda_{ZE}$, leading to the possibility of strong coupling. Figure \ref{Fig:f1}c shows $R_p$ for the combined FP cavity-ENZ mode structure (Ag-PMMA-CdO-Ag, hereafter ENZ-FP cavity; see inset), revealing substantially modified absorption features. A clear anti-crossing of resonances is observed near $\lambda_{ZE}$, characterized by a splitting of the overall dispersion into an upper and lower branch. This opens up a highly reflecting window around $\lambda_{ZE}$ as indicated by the arrow in Figure \ref{Fig:f1}c. The Rabi splitting ($\Omega_R$ = 55 meV) satisfies the criterion for strong coupling \cite{novotny2010strong, rider2021something}, which is given as $\Omega_R > (\gamma_{FP}+\gamma_{ENZ})/2$,  where $\gamma_{FP} \approx$ 12 meV and $\gamma_{ENZ} \approx$ 25 meV are full widths at half maximum (FWHM) of the bare FP and ENZ resonances at the crossing point, respectively.  
The corresponding reflectance maps of the cavities for \textit{s}-polarized light ($R_s$) are shown in Figure S2, where the Berreman mode cannot be excited and only the FP resonance is observed in the ENZ-FP cavity. This clearly indicates that ENZ mode excitation is an integral part of the strongly coupled interaction observed in $R_p$. Further, it is seen that the features in $R_s$ may be effectively employed as a reference against which to compare the signatures of SC in $R_p$. 

To further understand the origin of the splitting, the frequency-wavenumber dispersion relations of the upper and lower branches are modeled by the following expression \cite{runnerstrom2018polaritonic}:
\begin{equation}
	\omega^{\pm}(k) = \frac{\omega_{FP}(k)+\omega_{ENZ}(k)}{2} \pm \\ \frac{1}{2}\left[ \Omega_R^2+ \left( \omega_{FP}(k)-\omega_{ENZ}(k)\right)^2  \right] ^{1/2} 
	\label{eq:sc}
\end{equation}
where $\omega^{\pm}(k)$ denotes the dispersion relation of the upper (+) and lower (-) branches of the ENZ-FP cavity, $\omega_{FP}(k)$ is the dispersion relation of the bare FP cavity, $\omega_{ENZ}(k)$ is the dispersion relation of the bare ENZ mode, $k = k_0 \sin\theta$ is the in-plane wave number and $k_0$ is the incident free-space wave number. The details of the calculation of $\omega_{FP}(k)$ and $\omega_{ENZ}(k)$ are given in Section S3. The strong coupling  model in Equation \eqref{eq:sc} is presented in Figure \ref{Fig:f1}d. $\omega_{FP}(k)$ and $\omega_{ENZ}(k)$ are plotted as dashed curves, while the upper and lower branches $\omega^{\pm}(k)$ are plotted as solid curves, calculated using $\Omega_R$ estimated from Figure \ref{Fig:f1}c. For convenience, the point where the bare dispersions cross is referred to by ($\omega_{sc}$, $k_{sc}$). The model qualitatively reproduces the observed anti-crossing behavior and splitting around this point well. It also shows how the upper and lower branches asymptotically tend to the bare dispersions away from ($\omega_{sc}$, $k_{sc}$), validating the strongly coupled nature of interaction of the ENZ mode and the FP cavity at the crossing point. The anti-crossing behavior is further characterized in Supplementary Figure S4, which clearly shows mode splitting around $\lambda_{ZE}$ and support the strongly coupled interaction picture in the ENZ wavelength regime.

\subsection{Factors determining strong coupling}

In this section, the factors affecting the strength of SC are systematically investigated.
Strongly coupled systems have been classically described using a coupled harmonic oscillator model where the energy splitting depends on a coupling term in the coupled mode equations \cite{novotny2010strong}. In optical systems, this coupling can be ascribed intuitively to the spatial overlap between electric fields of the interacting resonances at ($\omega_{sc},k_{sc}$) \cite{schlather2013near}. In the case of the ENZ-FP cavity, this can be written as \cite{runnerstrom2018polaritonic}
\begin{equation}
	\Omega_R \propto \int_V \textbf{E}_{\text{ENZ}}(\textbf{r}).\textbf{E}_{\text{FP}}(\textbf{r})\text{d}V
	\label{eq:omega_r}
\end{equation} 
where $\textbf{E}_{\text{FP}}$ is the electric field associated with the FP cavity, $\textbf{E}_{\text{ENZ}}$ is the field of the ENZ mode, $\textbf{r}$ is the position vector, the integral is over the cavity volume $V$, and the fields are evaluated at ($\omega_{sc},k_{sc}$). The ratio $g$ of the frequency splitting $\Omega_R$ and the zero-epsilon frequency $\omega_{ZE}$,

\begin{equation}
	g = \frac{\Omega_R}{\omega_{ZE}}
\end{equation}
can be used to further quantify the coupling strength between the ENZ mode and the FP cavity \cite{manukyan2021dependence}.
Several approximations can be made to simplify the analysis of Equation \eqref{eq:omega_r}. First, the electric field of the ENZ mode is dominantly out-of-plane (along \textit{z} direction, see coordinate axes in Figure \ref{Fig:f1}a) \cite{campione2015theory}. Second, the out-of-plane ENZ mode fields in the ENZ-FP cavity are strongly confined to the interior of the ENZ layer, which results in the overlap integral being negligible outside the layer. Third, the field inside the ENZ layer is spatially uniform along \textit{z} \cite{campione2015theory}, allowing it to be taken outside the integral. These considerations allow Equation \eqref{eq:omega_r} to be simplified as 
\begin{equation}
	\Omega_R \;\; \propto \;\; E_{z,ENZ} \int_0^{d_{ENZ}} E_{z,FP}(z)dz
	\label{eq:omega_r1}
\end{equation}
where the subscript '\textit{z}' denotes the out-of-plane component of the electric fields and the integral is now limited to the thickness of the ENZ layer.

\begin{figure}
	\includegraphics[width=0.5\textwidth]{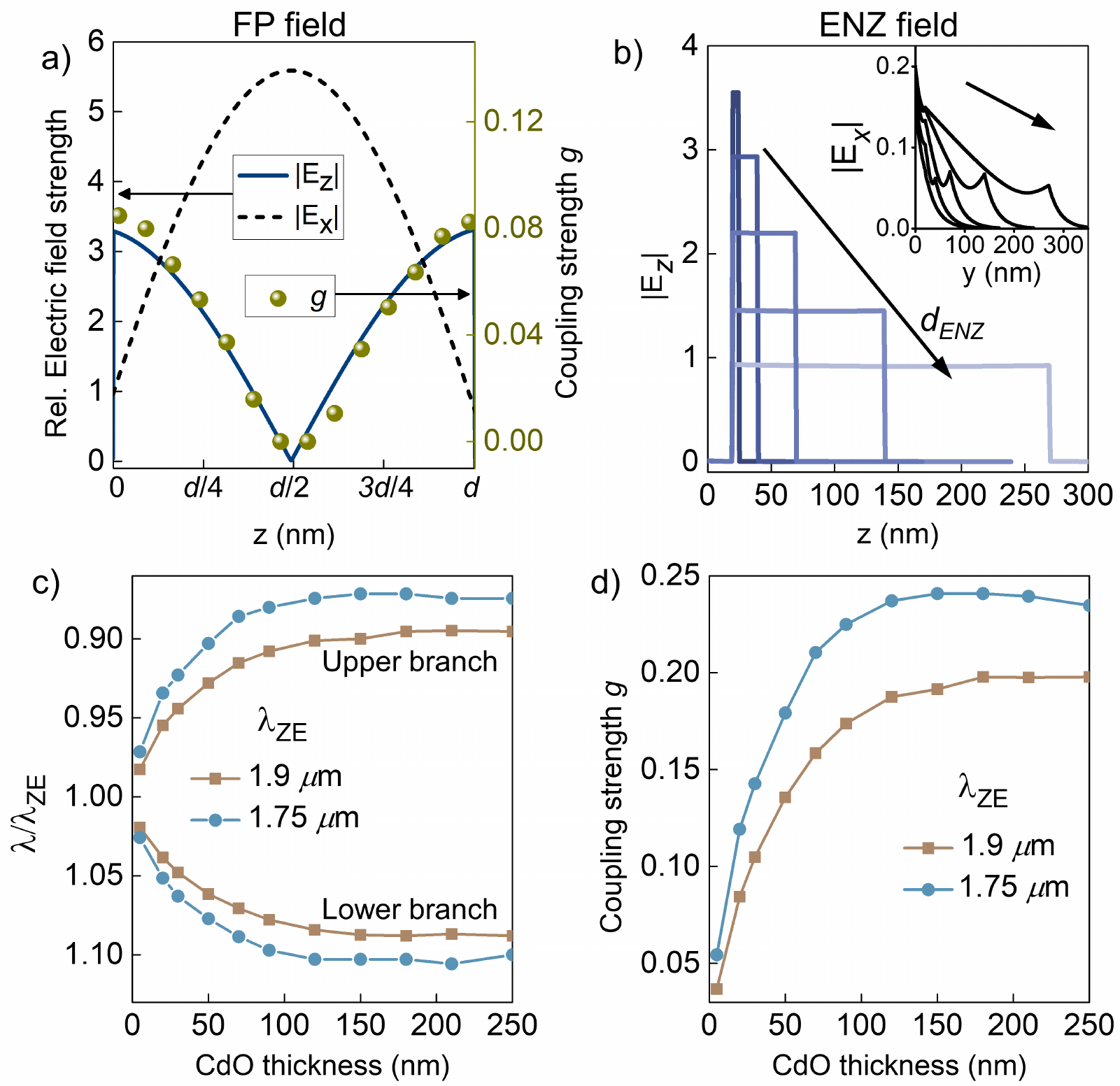}%
	\caption{(a) Left axis: Spatial variation of $E_{z,FP}$ and $E_{x,FP}$ field components (relative to the incident field strength) inside the bare FP cavity with vertical position ($z$) at the wavelength and angle where the bare dispersions cross. Right axis: The dependence of the coupling strength $g$ on the position of the ENZ layer relative to the FP cavity is co-plotted on the right axis. For this, $z$ denotes the position of the center of the ENZ layer with respect to the FP cavity.	(b) $E_{z,ENZ}$ inside the bare ENZ cavity for $d_{ENZ}$ varying from 5 to 250 nm at the wavelength and angle where the bare dispersions cross. Corresponding $E_{x,ENZ}$ plots are shown in the inset. Arrows indicate increasing $d_{ENZ}$. (c) Resonance wavelength splitting of the ENZ-FP cavity at ($\omega_{sc},k_{sc}$) as a function of varying $d_{ENZ}$ for $\lambda_{ZE}$ = 1.9, 1.75 $\mu$m. (d) Corresponding variation in $g$ with $d_{ENZ}$. \label{Fig:f3}}
\end{figure}

Based on Equation \eqref{eq:omega_r1}, the dependence of coupling strength on the FP cavity field is investigated first. Figure \ref{Fig:f3}a (left axis) plots the components of electric field as a function of vertical position inside the bare FP cavity. The fields are plotted for the FP cavity in Figure \ref{Fig:f1} ($d$ = 670 nm, $d_{top}$ = 20 nm), at the point ($\omega_{sc},k_{sc}$). The plot shows that while the in-plane component ($E_{x,FP}$) is minimum at the surface of the Ag mirrors, the out-of-plane component ($E_{z,FP}$) is maximum as a consequence of the Fresnel reflection phase imparted by the perfect electric conductor-like metal. Equation \eqref{eq:omega_r1} suggests that the strong coupling interaction, which is expected to be mediated by the \textit{z} components, will be maximized when the location of the ENZ layer is at the bottom (or top) of the ENZ-FP cavity (where $E_{z,FP}$ is maximum) and minimized when the ENZ layer is at the center of the cavity. To verify this, the coupling strength $g$ for the ENZ-FP cavity in Figure \ref{Fig:f1} is calculated as a function of the vertical position of the ENZ layer in the cavity and is plotted on the right axis of Figure \ref{Fig:f3}a. Here, the ENZ thickness is a constant and only its position within the cavity is varied.
It is evident that $g$ is maximum when the ENZ layer is at the surface of the Ag mirrors and is zero when it is placed at the center of the cavity, in close correlation with the strength of $E_z$ of the FP cavity. Note that this is in contrast to most observations in literature where the strong coupling is maximum when the active layers with dominantly in-plane fields or dipole moments are placed at the center of the FP cavity \cite{bhatt2021electromagnetic}.

Having identified the role of the out-of-plane FP field strength, the dependence of $\Omega_R$ on $d_{ENZ}$ and $E_{ENZ}$ are considered next. To illustrate this, the wavelength splitting of the upper and lower branches of the ENZ-FP cavity are calculated for different ENZ thicknesses. Figure S5 plots $R_p$ of the ENZ-FP cavity in Figure \ref{Fig:f1} for different ENZ thicknesses from 0 to 250 nm. From this, the wavelength splitting is calculated after identifying the respective cross-over points of the bare dispersions for each thickness. Figure \ref{Fig:f3}c plots the upper and lower polariton wavelengths (normalized to $\lambda_{ZE}$) against $d_{ENZ}$, calculated for $\lambda_{ZE}$ = 1750 and 1900 nm. The splitting is observed to initially increase with ENZ thickness but saturates around $d_{ENZ} \sim$ 100 nm for both values of $\lambda_{ZE}$. 
Figure \ref{Fig:f3}d further plots the variation of $g$ with $d_{ENZ}$, clearly showing the initial sharp increase in the coupling strength and its saturation at larger values of ENZ thickness for both the values of $\lambda_{ZE}$.
To understand this, two cases are analyzed here: \texttt{I}. $d_{ENZ}$ is a small fraction of the overall cavity thickness ($d_{ENZ} \ll d$) and \texttt{II}. $d_{ENZ} \sim d$. 
In case \texttt{I}, Equation \eqref{eq:omega_r1} can be further simplified by assuming that the spatial variation in $E_{FP}$ is negligible over the scale of the ENZ layer thickness, bringing it outside the integral. This yields \( \Omega_R \propto  E_{z,FP}E_{z,ENZ}d_{ENZ}\), which indicates that $\Omega_R$ increases in proportion to $d_{ENZ}$ when the ENZ thickness is low enough. More accurately, $\Omega_R$ is decided by the inter-relation between $d_{ENZ}$ and its mode field, $E_{z,ENZ}$. Figure \ref{Fig:f3}b plots $E_z$ inside the bare ENZ cavity by varying $d_{ENZ}$ from 5 nm to 250 nm, at ($\omega_{sc},k_{sc}$) where it crosses the FP dispersion of Figure \ref{Fig:f1}. The $E_z$ component is seen to be spatially uniform even for the thickest 250 nm CdO layer. The corresponding plots of $E_{x,ENZ}$ are shown in the inset, which are an order of magnitude weaker than $E_{z,ENZ}$, validating the assumptions stated earlier that the field in the ENZ mode is dominantly out-of-plane and spatially uniform in nature. Notably, $E_{z,ENZ}$ decreases as $d_{ENZ}$ increases, which means that the expected increase in coupling strength with $d_{ENZ}$ due to a larger interacting volume would be tempered by the decreasing ENZ mode field strength. Thus, the coupling strength will be determined by a trade-off between the larger interaction volume at large $d_{ENZ}$ and the stronger interacting electric field at small $d_{ENZ}$ in the ENZ layer. 

The initial increase in coupling strength in Figure \ref{Fig:f3}d follows from the expected dependence of $\Omega_R$ on $d_{ENZ}$ (larger interaction volume) discussed in case \texttt{I}. The increase is however sub-linear, pointing to the effect of decreasing $E_{z,ENZ}$ that reduces the interaction strength.
This does not, however, explain the saturation of $g$ at large $d_{ENZ}$, for which case \texttt{II} ($d_{ENZ} \sim d$) is considered. Here, it is evident that the variation in $E_{FP}$ over the scale of the ENZ layer thickness is no longer negligible. In fact, increasing the ENZ thickness can be imagined to be similar to adding ENZ layers sequentially towards the center of the FP cavity. As discussed earlier, this would contribute negligibly to the overall coupling strength due to the decay of $E_{z,FP}$ towards the center. Therefore, effectively only the ENZ layers near the surface of the Ag mirrors will be involved in SC, which qualitatively explains why $g$ saturates at large values of $d_{ENZ}$.

\subsection{Wide-angle polarizer design}

\begin{figure*}
	\includegraphics[width=\linewidth]{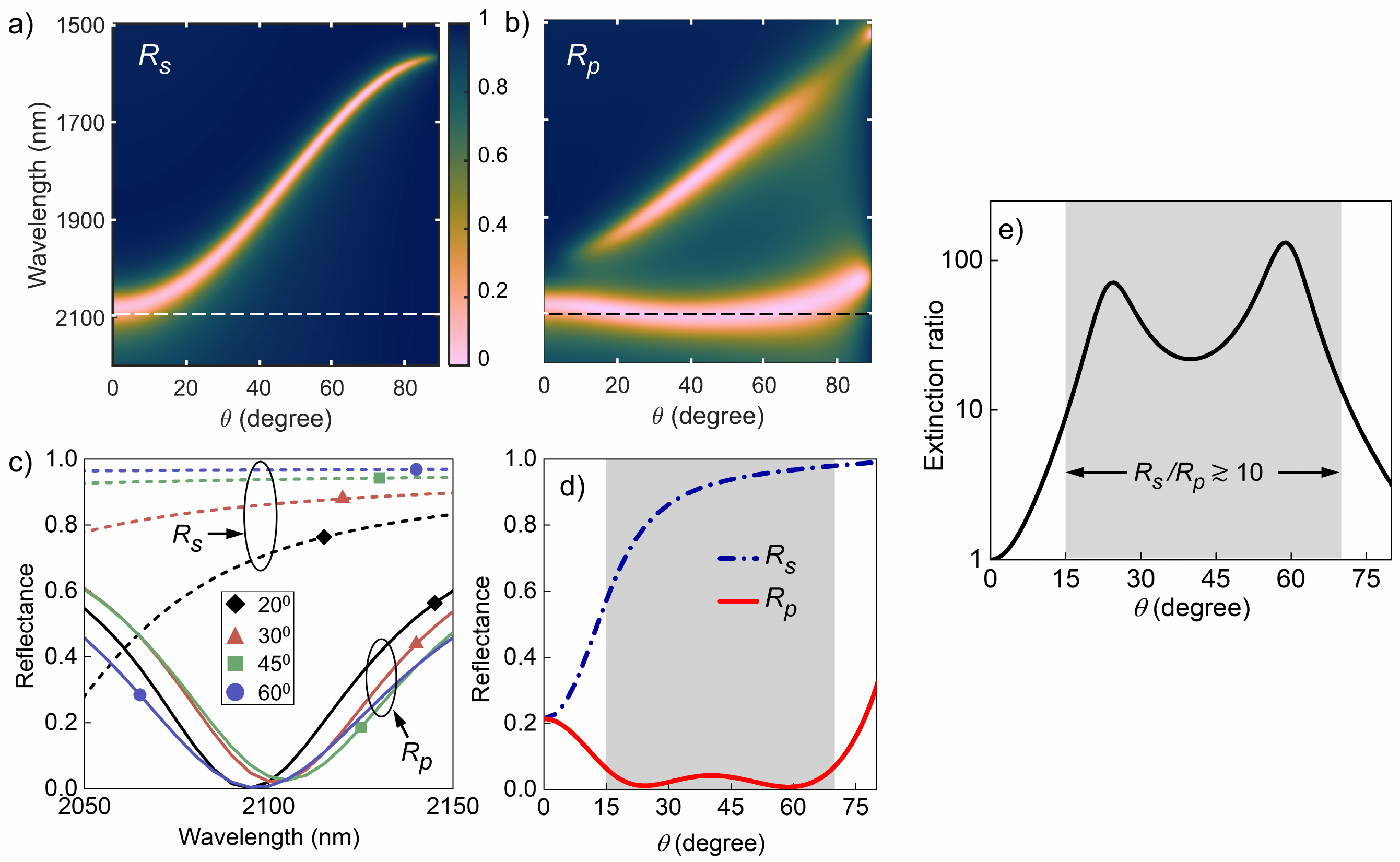}%
	\caption{(a,b) Calculated reflectance maps (a) $R_s$ and (b) $R_p$ of the Ag-PMMA-CdO-Ag cavity optimized as a wide-angle reflective \textit{s}-polarizer. The target wavelength $\lambda$ = 2100 nm is indicated by the horizontal lines. (c) Spectral variation of $R_s$ and $R_p$ at different incident angles around $\lambda$ = 2100 nm. (d) Angle-dependence of $R_s$ and $R_p$ plotted at $\lambda$ = 2100 nm. (e) $R_s/R_p$ at $\lambda$ = 2100 nm showing a reflectance contrast $>$ 10 for angles between 15 and 70 degrees.    
		\label{Fig:f4}}
\end{figure*}

Having identified the control parameters for strong coupling in an ENZ-FP cavity, in this section, the design of infrared wide-angle polarizers is demonstrated to highlight the extent of dispersion engineering possible in this system. Figure \ref{Fig:f4}a,b shows the reflectance maps $R_s$ and $R_p$ of an Ag-PMMA-CdO-Ag cavity designed to act as a wide-angle, reflective \textit{s}-polarizer at a target near-IR wavelength $\lambda$ = 2100 nm. The parameters of the cavity are $d$ = 522 nm, $d_{top}$ = 9 nm, $d_{ENZ}$ = 128 nm and $\lambda_{ZE}$ = 1.985 $\mu$m, as summarized in Table \ref{tab:Table1}. The numerical optimization process is outlined in Section S6.
In Figure \ref{Fig:f4}a, although $R_s$ is very low for near-normal incidence at $\lambda$ = 2100 nm, the strong angle-dependence of the bare FP dispersion ensures that the cavity strongly reflects \textit{s}-polarized light at oblique angles $\gtrsim$ 15$^0$. On the other hand, at this wavelength the cavity shows strong absorption of \textit{p}-polarized light for almost any angle of incidence (Figure \ref{Fig:f4}b), resulting in an extremely low $R_p$ over a wide range of $\theta$. This is achieved by engineering the SC in the ENZ-FP cavity so that its lower polariton branch lies near $\lambda$ = 2100 nm and possess an extremely flat dispersion, as evident in Figure \ref{Fig:f4}b.
This large contrast between $R_s$ and $R_p$ over a wide range of oblique angles effectively makes the ENZ-FP cavity a nearly omnidirectional \textit{s}-polarizer in reflection mode. To quantify the performance of the polarizer at and around its target wavelength, Figure \ref{Fig:f4}c shows the contrast between $R_s$ and $R_p$ around $\lambda$ = 2100 nm for four angles in the range 20 to 60 degrees. The minima of $R_p$ lie consistently close to zero near 2100 nm and show negligible spectral variation with $\theta$, while $R_s$ remains high throughout, showing the large and nearly angle-independent nature of the reflectance contrast. To further verify the extent of omnidirectionality, Figure \ref{Fig:f4}d plots the reflectance as a function of $\theta$ at 2100 nm. Away from normal incidence, R$_s$ steadily increases while R$_p$ decreases to nearly zero. In the shaded region where $\theta$ lies between 15$^0$ and 70$^0$, R$_s$ varies from $\approx$ 60\% to 98\% while R$_p$ lies below 7\% throughout this range, with an average value of 3\%. The extinction ratio, defined as the ratio of the power in orthogonal polarizations \cite{shahsafi2020infrared} (here $R_s/R_p$) is plotted in Figure \ref{Fig:f4}e. In the range $\theta$ = 15$^0$ - 70$^0$, the extinction ratio is at least 10 and goes up to as high as 10$^2$, showing that the designed ENZ-FP cavity can function as a polarizer with a wide working range of incident angles and high efficiency.

\begin{table*} []
	\centering
	\caption{Summary of numerically optimized parameters for wide-angle polarizers}
	\begin{tabular}{|c|c|c|}
		\hline
		Parameter 	& \textit{s}-polarizer	& \textit{p}-polarizer	\\ 
		\hline
		Target wavelength  & 2100 nm  & 4000 nm \\
		\hline
		Top metallic layer  & Ag, $d_{top}$ = 9 nm  & CdO, $d_{top}$ = 120 nm	\\
		\hline
		Dielectric layer  & PMMA, \textit{d} = 522 nm  & GST, \textit{d} = 107 nm	\\
		\hline
		CdO ENZ layer  &  $d_{ENZ}$ = 128 nm, $\lambda_{ZE}$ = 1.985 $\mu$m  & $d_{ENZ}$ = 118 nm, $\lambda_{ZE}$ = 3.98 $\mu$m	\\
		\hline
	\end{tabular}
	\label{tab:Table1}
\end{table*}

\begin{figure*}
	\includegraphics[width=0.65\textwidth]{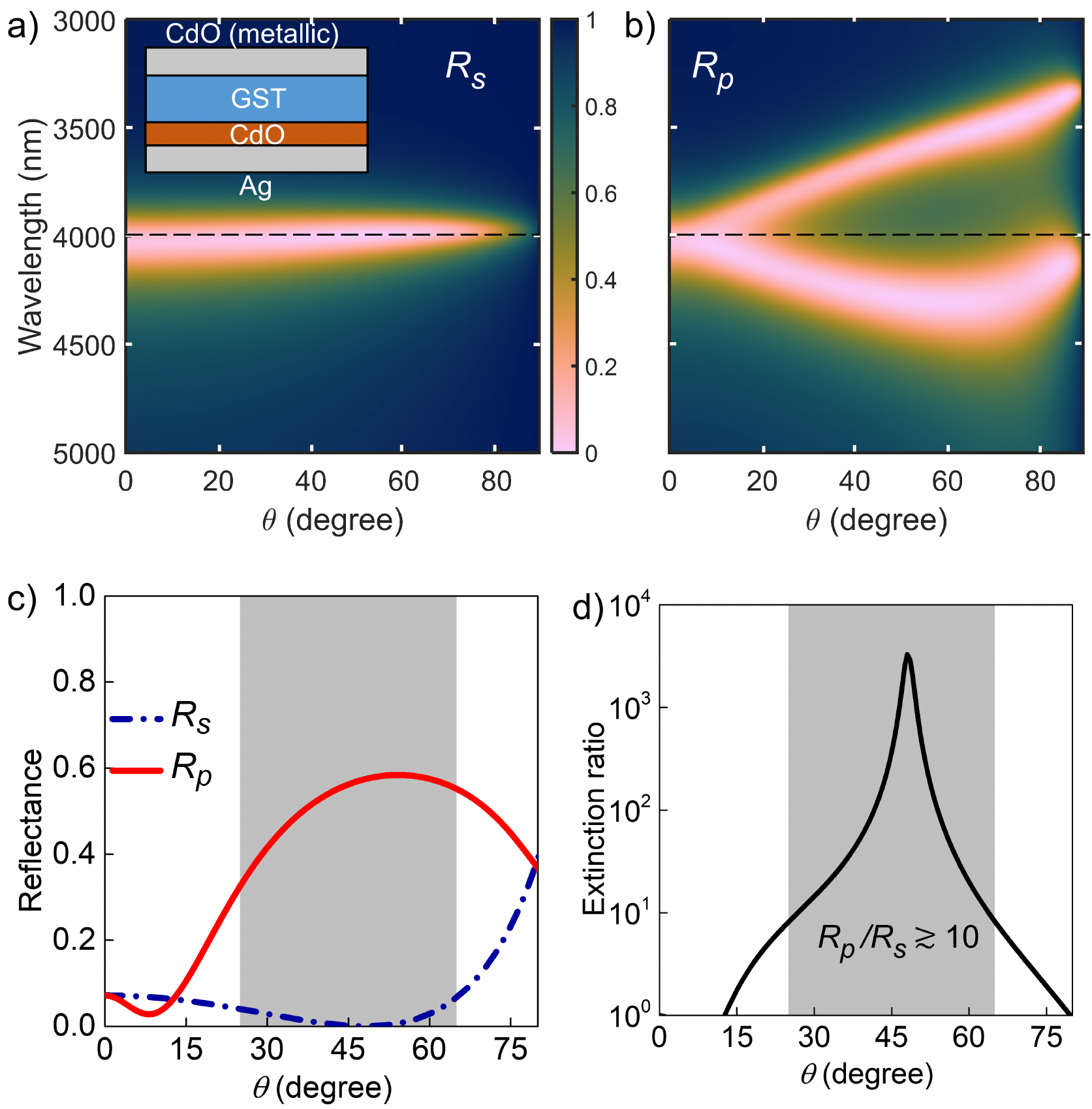}%
	\caption{(a,b) Calculated reflectance maps (a) $R_s$ and (b) $R_p$ of the mid-IR ENZ-FP cavity, shown schematically in the inset. The target wavelength $\lambda$ = 4000 nm is indicated by the horizontal lines. (c) Angle-dependence of $R_s$ and $R_p$ plotted at $\lambda$ = 4000 nm. (d) $R_p/R_s$ at $\lambda$ = 4000 nm showing an extinction ratio $\sim$ 10 to 10$^3$ for angles between 25 and 65 degrees.  \label{Fig:f5}}
\end{figure*}

Following this, a reflective \textit{p}-polarizer with a wide angular range is also realized based on SC in the ENZ-FP cavity. Such a device would have omnidirectional absorption of \textit{s}-polarized light and high reflectance for \textit{p}-polarized light, requiring opposite results to that in Figure \ref{Fig:f4}. Polarizer schemes based on thin film polaritonic absorbers usually involve absorption of the \textit{p}-component since surface polaritons are excited only by transverse magnetic fields \cite{vassant2012berreman}. In light of this, a polarizer that can reflect \textit{p}- and absorb \textit{s}- polarized light over wide angles would be an important advancement. This requires an all-angle absorber for \textit{s}-polarization with a very flat dispersion at the target wavelength. However, $R_s$, which is determined by the bare FP cavity resonance, shows a prominently dispersive behavior. For example, in Figure \ref{Fig:f4}a the resonance wavelength varies by $\sim$ 500 nm in the near-IR as $\theta$ is varied from 0 to 90$^0$. To overcome this dispersive nature of the FP resonance and obtain angle-independent absorption, crystalline germanium antimony telluride (GST) having an extremely high refractive index ($n>6$) with a simultaneously low loss ($k \approx$ 0.05) in the mid-IR window from 3 to 5 $\mu$m \cite{michel2013using} is used as the cavity dielectric. Section S7 further discusses the dependence of the FP resonance on the dielectric refractive index.
Due to operation in the mid-IR window (3 - 5 $\mu$m), the Ag top layer is replaced with a metallic CdO layer with $\lambda_{ZE}$ = 2.1 $\mu$m. The metallic CdO layer only plays the role of mirror in the FP cavity, which is used because Ag becomes too reflective to allow light to pass into the cavity.
Setting a target operation wavelength of $\lambda$ = 4 $\mu$m for the polarizer, Figure \ref{Fig:f5}a shows $R_s$ for an ENZ-FP cavity (shown schematically in the inset), revealing an extremely flat dispersion independent of $\theta$ at the target wavelength. The numerically optimized parameters for this ENZ-FP cavity are $d_{top}$ = 120 nm, $d$ = 107 nm, $d_{enz}$ = 118 nm and $\lambda_{ZE}$ = 3.98 $\mu$m (Table \ref{tab:Table1}). The corresponding $R_p$ map is shown in Figure \ref{Fig:f5}b, showing the upper and lower polaritonic branches of the ENZ-FP cavity. The resulting strong coupling ensures a large splitting between the upper and lower branches at oblique incidence, which opens up a reflecting spectral window in $R_p$ at the target wavelength over a wide angular range (Figure \ref{Fig:f5}b). It is interesting to note that the numerically optimized ENZ wavelength of $\lambda_{ZE}$ = 3.98 $\mu$m is spectrally coincident with the FP resonance in Figure \ref{Fig:f5}a, which facilitates the opening of the reflecting window at the target wavelength. To demonstrate the wide angular range of the reflectance contrast, Figure \ref{Fig:f5}c plots $R_s$ and $R_p$ at $\lambda$ = 4 $\mu$m as a function of $\theta$. A large contrast in reflectance at angles between 25$^0$ and 65$^0$ in the shaded region is evident where R$_p$ varies from $\approx$ 35\% to 60\% while R$_s$ lies below 7\% throughout this range, with an average value of 2\%.
Further, the extinction ratio (here $R_p/R_s$) plotted in Figure \ref{Fig:f5}d demonstrates that the ratio is greater than 10 within the shaded region between 25$^0$ and 65$^0$, reaching a maximum value $>$ 3000 around 50$^0$ where $R_s$ goes to zero. Thus, engineering the ENZ-FP dispersion via SC is shown to give a high-efficiency, wide-angle \textit{s-} polarizer in the mid-IR. 

It is worth mentioning here that these cavities also hold promise in thermal photonics applications. For example, Figure \ref{Fig:f5}a shows an omnidirectional, wavelength-selective perfect absorber, which is also a polarized, selective thermal emitter according to Kirchhoff's law \cite{picardi2023dynamic}. In fact, both the structures shown in Figure \ref{Fig:f4} and Figure \ref{Fig:f5} are highly polarized, wavelength-selective thermal emitters. Significantly, previous demonstrations of wide-angle, selective emitters have utilized photonic crystals and nanostructures, which present a much larger scale of fabrication complexity compared to the multilayer structures here \cite{giteau2022active}.
Another attractive feature of these cavities is their tunability. Although the polarizers are designed here at only a particular infrared wavelength, there is nothing special about the demonstrated operation wavelengths of $\lambda$ = 2100 nm and 4000 nm. To illustrate this, note that apart from the geometric parameters of the cavity, the dielectric parameters of all the materials employed here are also highly tunable. For instance, the ENZ wavelength of transparent conducting oxides such as ITO and doped CdO can be tuned over a wide range of wavelengths \cite{johns2020epsilon,runnerstrom2018polaritonic}. This critical property is what allows $\lambda_{ZE}$ to be included as an optimization parameter in our calculations. Furthermore, GST is a phase change material showing non-volatile switching of its optical response in the visible and infrared regions \cite{cueff2021reconfigurable,wuttig2017phase,parra2021toward}. Thus, apart from the static tunability, this opens up the exciting possibility of dynamically tuning the strongly coupled cavity interactions \cite{picardi2023dynamic}.

\subsection{Dynamic multi-level resonances in ENZ-FP cavity}

\begin{figure}
	\includegraphics[width=\textwidth]{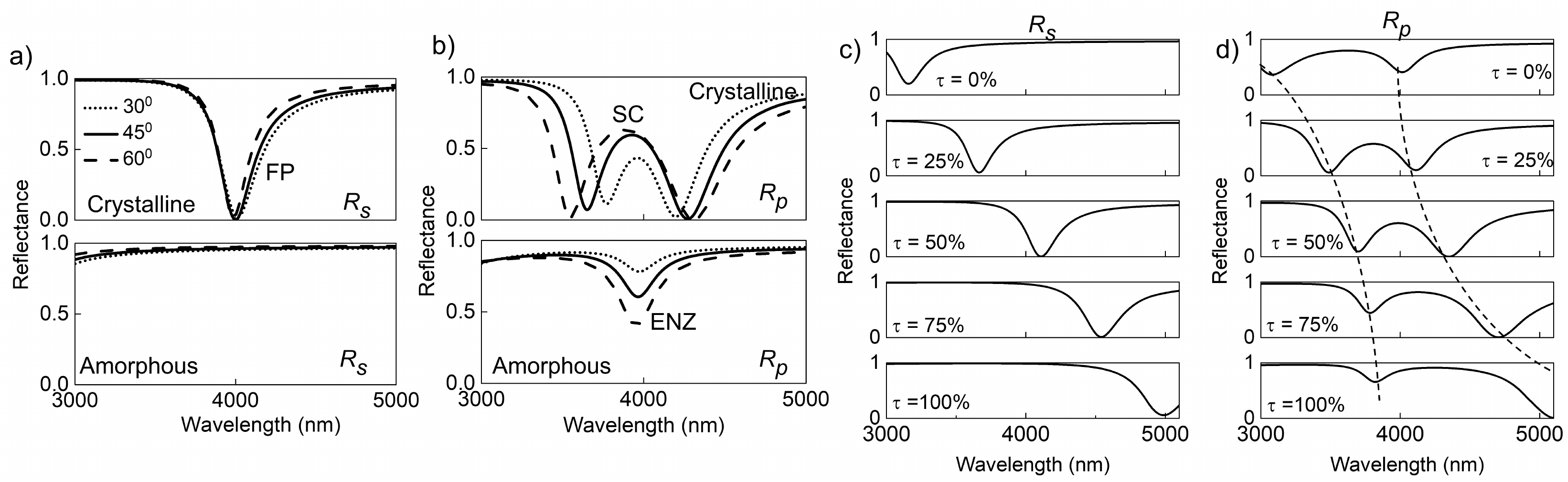}%
	\caption{\label{Fig:f6}
		(a) Wavelength dependence of $R_s$ of the mid-IR ENZ-FP cavity for crystalline GST (top) and amorphous GST (bottom). The dip in $R_s$ is due to the FP resonance. (b) Corresponding plots of $R_p$ for the ENZ-FP cavity. The top panel shows strong coupling (SC) while the dip in the bottom panel is due to the ENZ mode. The legend showing angle of incidence is common to (a) and (b).
		Reflectance spectra (c) $R_s$ and (d) $R_p$ of the mid-IR ENZ-FP cavity at an angle of incidence 45$^0$ for different values of the crystallization fraction ($\tau$) of GST. Dashed lines in (d) indicate the avoided crossing of resonances.}
\end{figure}

In this section, the potential for active tunability of the ENZ-FP cavity by exploiting phase change in GST is explored. Figure \ref{Fig:f6}a compares the wavelength dependence of $R_s$ of the mid-IR ENZ-FP cavity for three values of $\theta$ when GST is in its crystalline phase (top panel) and amorphous phase (bottom). A remarkable variation in $R_s$ is evident with the cavity changing from a nearly omnidirectional perfect absorber at $\lambda$ = 4 $\mu$m to a perfect reflector of \textit{s}-polarized light on switching the phase of GST. The underlying reason here is that the FP resonance at 4 $\mu$m for crystalline GST has shifted to much lower wavelengths due to the lower refractive index of GST in its amorphous phase ($n \approx$ 3.5) \cite{michel2013using}. Figure \ref{Fig:f6}b shows the corresponding plots of $R_p$. Here, the strongly coupled resonances in the crystalline phase are evident as dual reflectance dips in the top panel. However, the dual resonances disappear when GST changes to its amorphous phase leaving a single resonance at the ENZ wavelength of CdO. This is because the FP resonance in the amorphous phase is spectrally separated from the ENZ mode, and the only response of the ENZ-FP cavity for \textit{p}-polarized light in the 3 - 5 $\mu$m range is the absorption of the ENZ mode near 4 $\mu$m, as seen in the bottom panel of Figure \ref{Fig:f6}b. 

This active tunability of the FP resonance in the ENZ-FP cavity opens up exciting possibilities for dynamic, multi-level tuning of the polariton dispersions. Particularly, the multi-state switchability of GST's phase has been shown to allow fine control over its optical response, owing to partial crystallization that leads to intermediate levels of refractive index between the crystalline and amorphous phases \cite{cueff2021reconfigurable}. Assuming that the refractive index of GST varies linearly between its two phases as a function of crystallization fraction ($\tau$), the mid-IR response of the ENZ-FP cavity as a function of $\tau$ is investigated. Figure \ref{Fig:f6}c plots the wavelength dependence of $R_s$ of the mid-IR ENZ-FP cavity at $\theta$ = 45$^0$ varying $\tau$ from 0\% (top) to 100\% (bottom). The plot indicates that by varying the refractive index of GST, a systematic, quasi-continuous tunability of the FP resonance condition is achieved.
The thickness of GST here is set to $d$ = 180 nm such that the resonances span the entire wavelength range from 3 to 5 $\mu$m as $\tau$ is varied. Figure \ref{Fig:f6}d shows the corresponding plots of $R_p$. At $\tau$ = 0 and 100\%, the FP resonances are far away from the ENZ mode at 4 $\mu$m, leading to two spectrally distant resonant features corresponding to the uncoupled ENZ and FP resonances. At intermediate values of $\tau$, the bare FP resonance approaches $\lambda_{ZE}$ resulting in distinct, $\tau$-dependent, strongly coupled ENZ-FP resonances typified by the avoided crossing of dispersions around 4 $\mu$m. Dashed lines are used to highlight this avoided crossing in Figure \ref{Fig:f6}d.
The multilevel resonances that can be finely controlled by the crystallization fraction leads to multiple-state dynamic switchability of the strong coupling interaction, i.e. not only can the strong coupling be switched between 'ON' and 'OFF', it can have continuously tuned states lying between these 'ON' and 'OFF' states. Such multi-level crystallization of GST may be experimentally realized using heating stages, electrical signals or optical pumping \cite{wuttig2017phase,parra2021toward}. Practically, studies have reported enough control over $\tau$ to realize over 100 distinct states, each of which correspond to different, non-volatile states of crystallization \cite{cueff2021reconfigurable}. This allows one to envision applications that improve from binary to gray scale functions, especially in continuous wave-front shaping in metasurfaces and improved dynamic control of spatial light modulators \cite{picardi2023dynamic}. Coupled with the unprecedented tunability of the polariton dispersion demonstrated here, these will be useful in the design of novel optical devices and advanced functionalities. 

\section{Conclusions} 	

In conclusion, strong coupling of optical resonances in a planar, multi-layer system of coupled ENZ and FP cavities is demonstrated. 
The work exploits the unique optical properties of the interacting resonances to present a simple geometric structure where ENZ light-matter interactions can be easily tailored, with the results demonstrating an unprecedented control over the ENZ mode dispersion.
The coupling strength, quantified by $\Omega_R$ and the ratio $g$, varies strongly with the thickness and field enhancement of the ENZ layer, as well as its position within the cavity. An analytical approach to elucidate these dependencies is presented by estimating field overlap of the interacting modes, which accurately predicts the trends in the variation of $\Omega_R$. The model is further validated by numerical calculations of field distribution and splitting in the system. $\Omega_R$ is shown to reach values as large as 20\% of the operating frequency, indicating unique and efficient mode coupling in a simple, planar structure. Through numerical optimization of the cavity geometry and leveraging the spectral tunability of the ENZ regime in doped CdO, the potential for extreme dispersion engineering in the system is demonstrated. 
Remarkably, the cavity can be designed as nearly-omnidirectional, wavelength selective polarizers, for both \textit{s} and \textit{p} polarizations. The tunable optical properties of the ENZ layer and GST dielectric layer open up the possibility of tuning the response over a wide spectral range by suitable material choice and geometry optimization as demonstrated here. In particular, the partial crystallization of GST allows quasi-continuous tuning of the cavity, providing an excellent handle for dynamic and non-volatile modulation of the SC dispersion through multiple resonant levels. The results presented here shed light on controlling and engineering ENZ light-matter interactions through coherent processes and are promising for the development of IR optical components e.g., thermal emitters \cite{xu2021broadband, johns2022tailoring} with active and tunable functionalities. 

\section*{\label{Methods} Methods}

\paragraph*{Material properties}
The Drude model is used to calculate the frequency-dependent permittivity $\epsilon(\omega)$ of Ag and CdO, given by 
\begin{equation}
	\epsilon(\omega) = \epsilon_{\infty} - \frac{\omega_p^2}{\omega^2+i\gamma\omega}
\end{equation}
where $\epsilon_{\infty}$ is the high-frequency permittivity, $\omega_p$ is the plasma frequency and $\gamma$ is the scattering rate. 
The Drude model parameters for Ag are $\epsilon_{\infty}$ = 5, $\omega_p$ = 8.9 eV and $\gamma$ = 0.039 eV \cite{yang2015optical}, and the fixed parameters in the Drude model for CdO are $\epsilon_{\infty}$ = 5.3 and $\gamma$ = 2.8 $\times$ 10$^{13}$ rad/s \cite{runnerstrom2018polaritonic} while $\omega_p$ is varied. 
In the low loss case ($\gamma \ll$ $\omega_p$), the real part of permittivity ($\epsilon'$) becomes zero at the zero-epsilon frequency $\omega_{ZE} = \omega_p/\sqrt{\epsilon_{\infty}}$. The corresponding wavelength is denoted as $\lambda_{ZE}$, which lies in the UV for Ag and can be tuned in the near to mid-IR range for CdO \cite{runnerstrom2018polaritonic}. In this work, CdO is considered throughout as the ENZ material with its $\lambda_{ZE}$ as a tunable parameter. 

The cavity dielectric materials used are either PMMA, having a refractive index $n_d$ = 1.47 at near-IR wavelengths \cite{kocer2015reduced}, or GST, which is a phase change material whose refractive index varies depending on its phases (amorphous or crystalline). 
In the mid-IR wavelength range where the properties of a GST-integrated FP cavity are analyzed, the complex refractive index varies approximately from a value of 6 in its crystalline state to 3.5 in its amorphous state, with the corresponding loss changing from $\approx$ 0.05 to practically 0 \cite{michel2013using}. The permittivity or refractive index plots are given in Section S1 in Supplementary Materials.

\paragraph*{Transfer matrix calculations}
A custom written transfer matrix method (TMM) code is used to calculate the reflectance and electric field distribution in the multilayers investigated here. The TMM calculations are described in Section S8 in Supplementary Materials.

\section*{Acknowledgements}
BJ thanks Dr. Jino George (Molecular Strong Coupling group, IISER Mohali) for his helpful comments, discussions, and encouragement.

\section*{Funding}
  BJ acknowledges IISER Mohali for Institute Postdoctoral fellowship. 

\bibliographystyle{unsrtnat}
\bibliography{enz_fp_refs}

\end{document}


\raggedbottom

\title{{\Large Supplementary Material for} \\ Dispersion engineering of infrared epsilon-near-zero modes by strong coupling to optical cavities}
	
\author{Ben Johns}
\email[]{benjohns@iisermohali.ac.in}
\affiliation{%
	Department of Chemical Sciences, Indian Institute of Science Education and Research, Mohali, India, 140306
}%

\date{30 March, 2023}
	
\maketitle

\makeatletter 
\renewcommand{\thefigure}{S\@arabic\c@figure}
\makeatother
\renewcommand{\thesection}{S\arabic{section}}

\section{\label{Materials} Optical constants}

\begin{figure*}[h]
	\centering
	\includegraphics[width=0.6\linewidth]{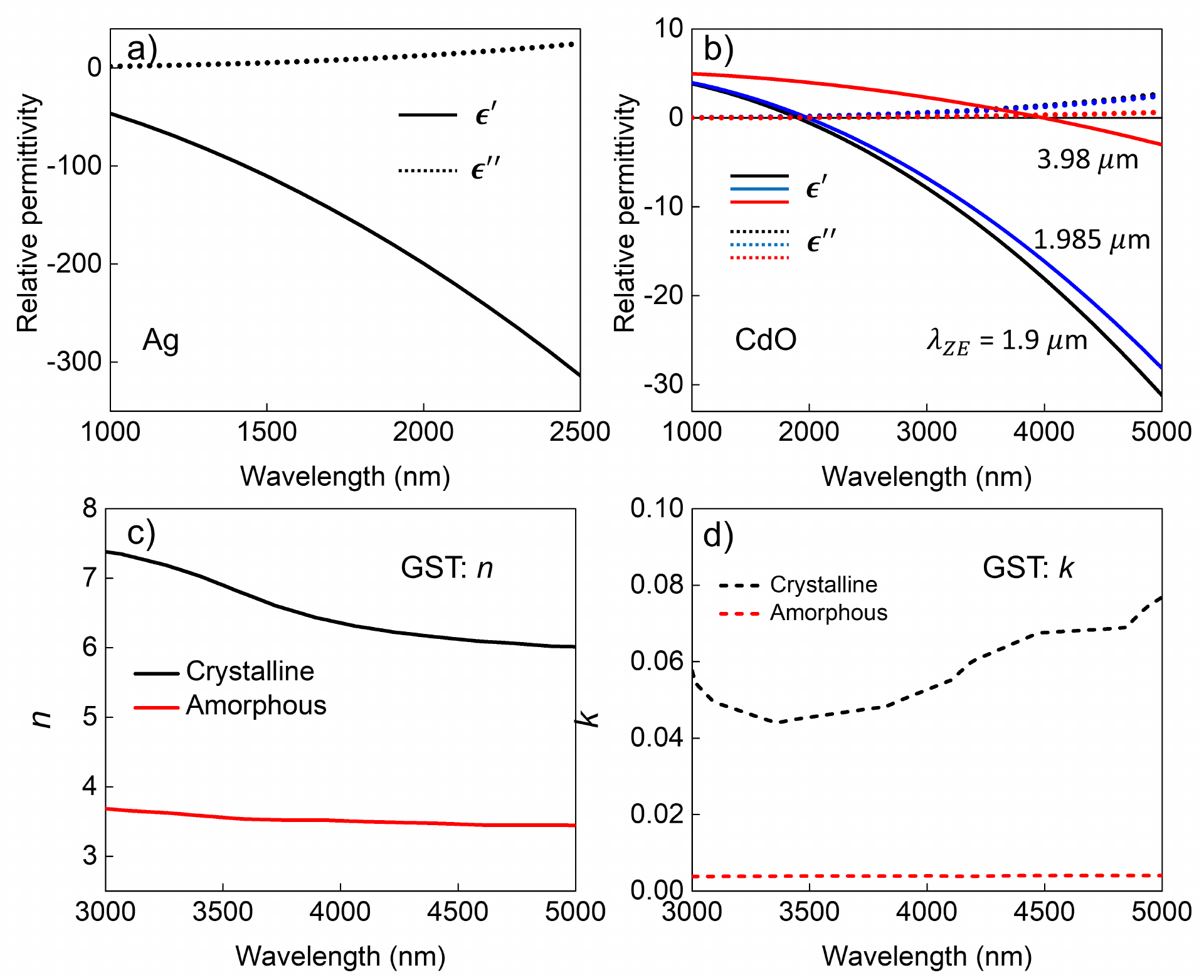}%
	\caption{(a) Permittivity of Ag, (b) permittivity of CdO with different values of $\lambda_{ZE}$, (c) real and (d) imaginary refractive index of GST in its two phases. \label{Fig:0}}
\end{figure*}

Figure \ref{Fig:0}a,b plots the real and imaginary permittivity of Ag and CdO. The permittivity of CdO for three values of $\lambda_{ZE}$ used in this work are shown as labeled in Figure \ref{Fig:0}b. The real ($n$) and imaginary ($k$) parts of GST refractive index in the crystalline and amorphous phases are plotted in Figure \ref{Fig:0}c,d \cite{michel2013using}.

\newpage
\section{\label{} $R_s$ maps of ENZ-FP cavity}

\begin{figure*}[h]
	\includegraphics[width=\textwidth]{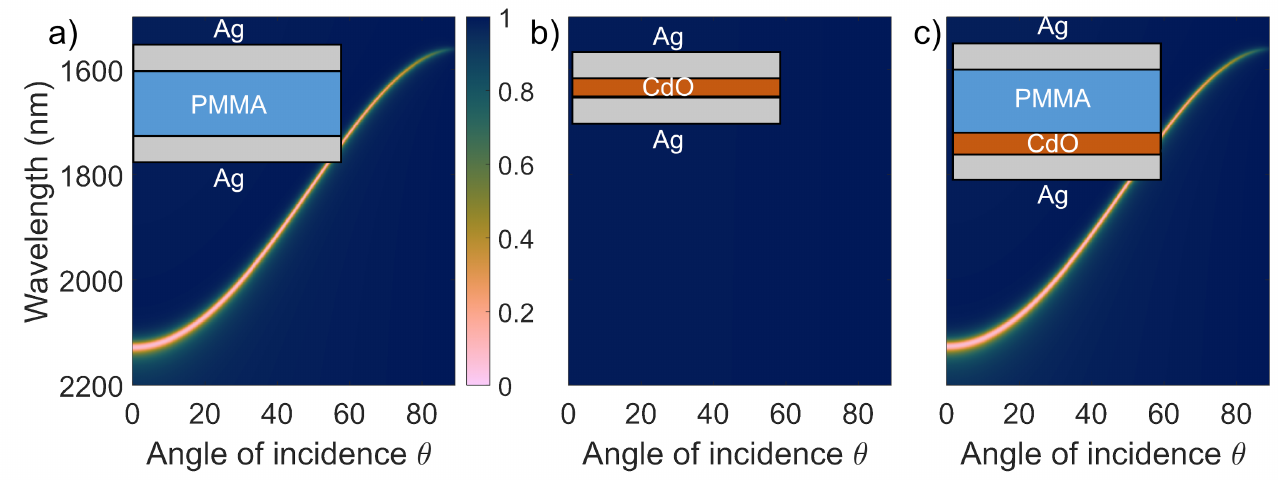}%
	\caption{$R_s$ maps of (a) FP cavity (b) ENZ mode structure and (c) ENZ-FP cavity, corresponding to the $R_p$ maps shown in Figure 1 in main text. All parameters are the same as in Figure 1. The absence any ENZ mode excitation in (b) implies that the ENZ-FP cavity response to s-polarized light in (c) is only due to the FP resonance.}\label{Fig:1}
\end{figure*}

\section{\label{} Bare dispersion relations}
\subsection*{Dispersion of FP cavity}

\begin{figure*}
	\includegraphics[width=\linewidth]{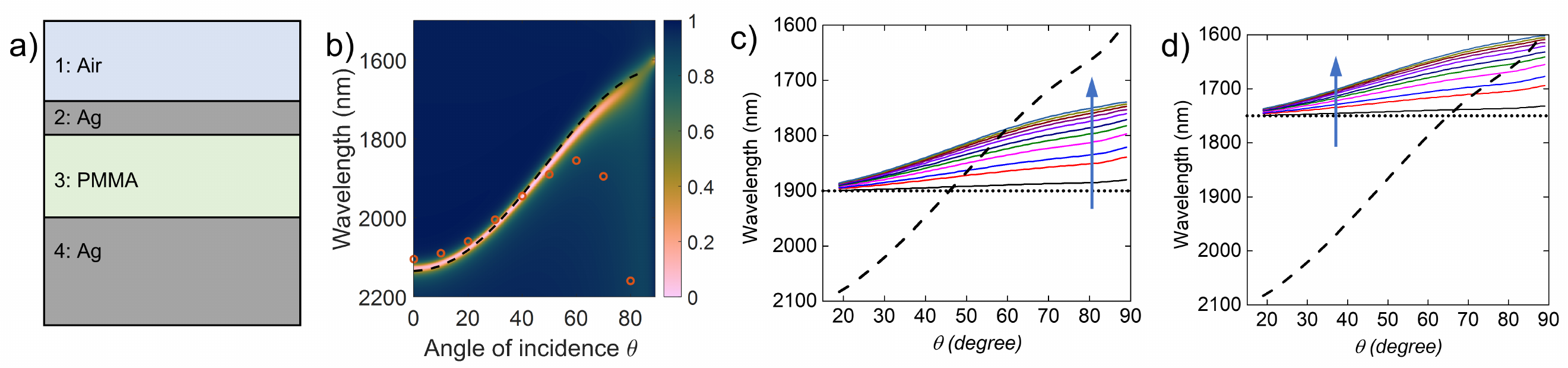}
	\caption{(a) Schematic of 4-layer structure with air above the FP cavity. (b) Dispersion relation of FP cavity plotted as dashed black curve, overlaid on the $R_p$ map from Figure 1a of main text. The Fabry-Perot resonance condition is identified by the red circles. (c,d) Dispersion relation of Ag-CdO-Ag ENZ mode structure for $\lambda_{ZE}$ = 1900 nm and 1750 nm, respectively. $d_{ENZ}$ is varied from 5 nm to 250 nm in each plot where the direction of increasing thickness is indicated by arrows. For comparison, the dashed curve shows the dispersion relation of FP cavity. \label{Fig:Sch1}}
\end{figure*}

Due to imperfect reflection from the Ag mirrors, the FP cavity resonance condition is shifted slightly from that of a conventional Fabry-Perot resonator, which occurs when there is a 2$\pi$ phase build-up over a round trip inside the cavity \cite{li2015large}. To accurately model the dispersion of the FP cavity, the condition for minimizing the reflectance of the layered structure in Figure \ref{Fig:Sch1}a is calculated. The reflection coefficient of the four layer structure can be written as 

\begin{equation}
		r_{1234} = \frac{r_{12} + r_{234}e^{2i\delta_2} }{1 + r_{12}r_{234}e^{2i\delta_2}}; 
		\label{eq:r1234}
\end{equation}

where \( r_{234} = \frac{r_{23} + r_{34}e^{2i\delta_3} }{1 + r_{23}r_{34}e^{2i\delta_3}}  \), $r_{ij}$ is the Fresnel reflection coefficient for media $i,j$, $\delta_i = k_{zi}d$ and $k_{zi}$ is the normal wave vector component in medium $i$. The phase condition to minimize the numerator in Equation \ref{eq:r1234} becomes 
\begin{equation}
	\phi_{12} = \phi_{234} + 2\, Re(\delta_2) + (2m+1)\pi 
	\label{phaseeqn1}
\end{equation}  
where $\phi$ denotes the phase. Assuming low phase accumulation in the thin Ag layer i.e. $Re(\delta_2) \rightarrow 0$, the phase condition becomes 	
\begin{equation}
	\phi_{12} = \phi_{234} + (2m+1)\pi 
	\label{phaseeqn2}
\end{equation} 

Numerical solution of Equation \ref{phaseeqn2} is used to obtain the dispersion relation of the FP cavity, plotted as the dashed black curve in Figure \ref{Fig:Sch1}b, overlaid on the reflectance map of \textit{p}-polarized light. The thicknesses are $d_{top}$ = 20 nm, $d$ = 670 nm, corresponding to Figure 1a in the main text. The $2\pi$ phase accumulation condition also is plotted as red circles in Figure \ref{Fig:Sch1}b, showing that at low angles, the cavity may be well approximated as a Fabry-Perot resonator. At larger angles, the deviations become more significant and the Fabry-Perot model is not well suited to describe the cavity at all angles of incidence.

\subsection*{Dispersion of ENZ mode}

Figure \ref{Fig:Sch1}c,d plots the dispersion relation of ENZ modes (Ag-CdO-Ag) obtained from the corresponding numerical reflectance spectra, for $\lambda_{ZE}$ = 1900 and 1750 nm, respectively. The thickness of CdO is varied from 5 nm to 250 nm with $d_{top}$ = 20 nm. The blue-shifting of ENZ modes with thickness at larger angles (shown by arrows) is a well known feature that has been described in literature \cite{passler2019second}. The FP cavity dispersion from Figure \ref{Fig:Sch1}b is also shown here to highlight the crossing of the FP and ENZ dispersions for both the $\lambda_{ZE}$ = 1900 and 1750 nm cases. The ENZ dispersions lie within $\sim$ 100 nm range below $\lambda_{ZE}$, with the lowest thicknesses having a rather flat dispersion with $\omega_{ENZ}(k) \approx \omega_{ZE}$ (horizontal dotted lines in Figure \ref{Fig:Sch1}c,d). The dispersion of the ENZ mode in Figure \textbf{1b} of the main text for $d_{enz}$ = 20 nm is therefore approximated to be $\omega_{ENZ}(k) = \omega_{ZE}$ for the coupled dispersion calculations.

\section{\label{} Anti-crossing of dispersions around the ENZ wavelength}

\begin{figure}[h]
	\includegraphics[width=0.45\textwidth]{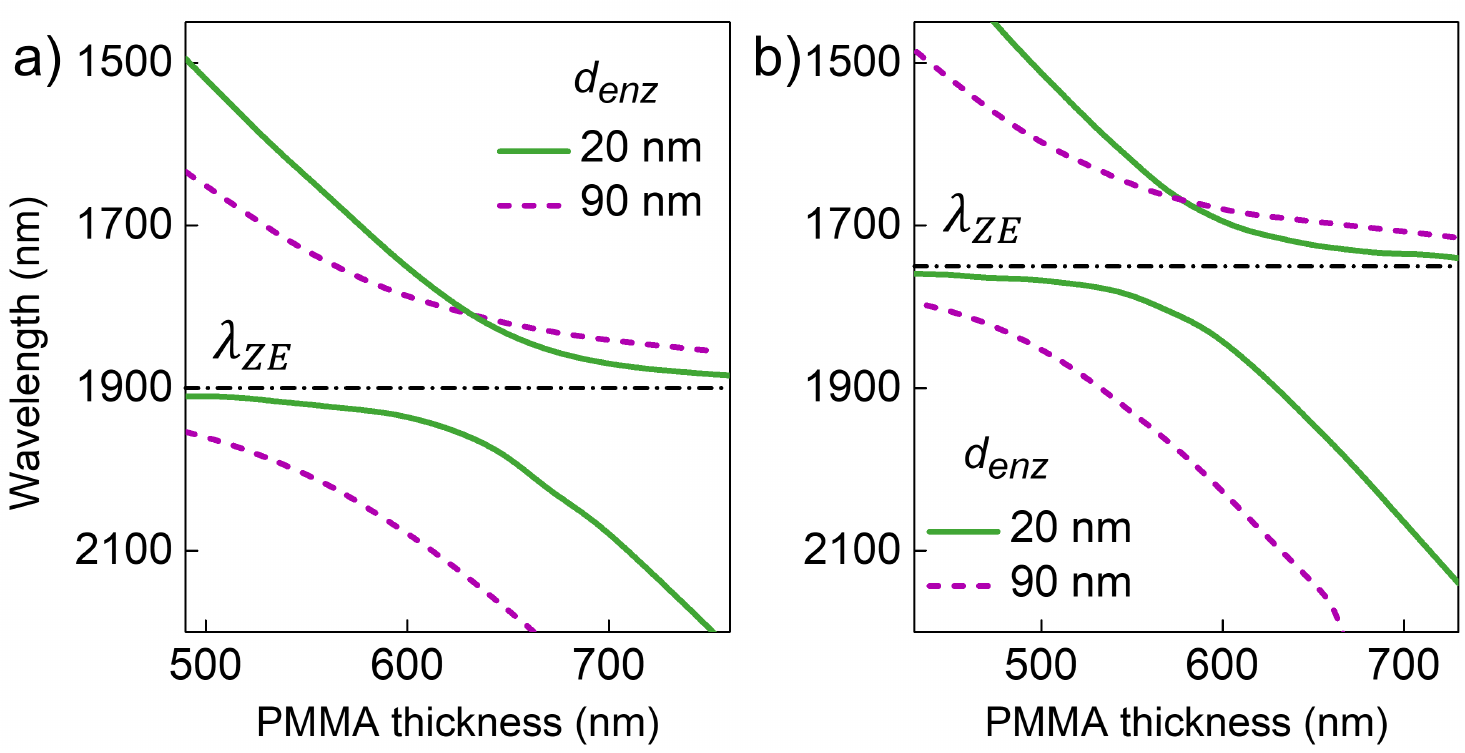}%
	\caption{Wavelength of the upper and lower branches of the ENZ-FP cavity at $\theta$ = 45$^0$ as a function of PMMA thickness, plotted for $d_{ENZ}$ = 20, 90 nm and (a) $\lambda_{ZE}$ = 1900 nm and (b) $\lambda_{ZE}$ = 1750 nm. Horizontal lines show $\lambda_{ZE}$, confirming that mode splitting and avoided crossing of the dispersions occur around the ENZ wavelength. \label{Fig:f2}}
\end{figure}

Figure \ref{Fig:f2} shows the dependence of the upper and lower branch wavelengths of the ENZ-FP cavity on PMMA thickness. The resonance wavelengths are calculated for two values of the ENZ thickness, $d_{ENZ}$ = 20 nm and 90 nm at a fixed $\theta$ = 45$^0$.
Figure \ref{Fig:f2}a shows the variation when the ENZ wavelength $\lambda_{ZE}$ is fixed at 1900 nm and Figure \ref{Fig:f2}b shows the plots for $\lambda_{ZE}$ = 1750 nm.
The mode splitting and anti-crossing behavior of the resonance wavelengths evident around $\lambda_{ZE}$ in both Figure \ref{Fig:f2}a and \ref{Fig:f2}b provide clear evidence for the strongly coupled interaction of the ENZ-FP cavity around $\lambda_{ZE}$. 

\newpage
\section{\label{}Dependence on ENZ thickness}

\begin{figure*}[h]
	\includegraphics[width=\linewidth]{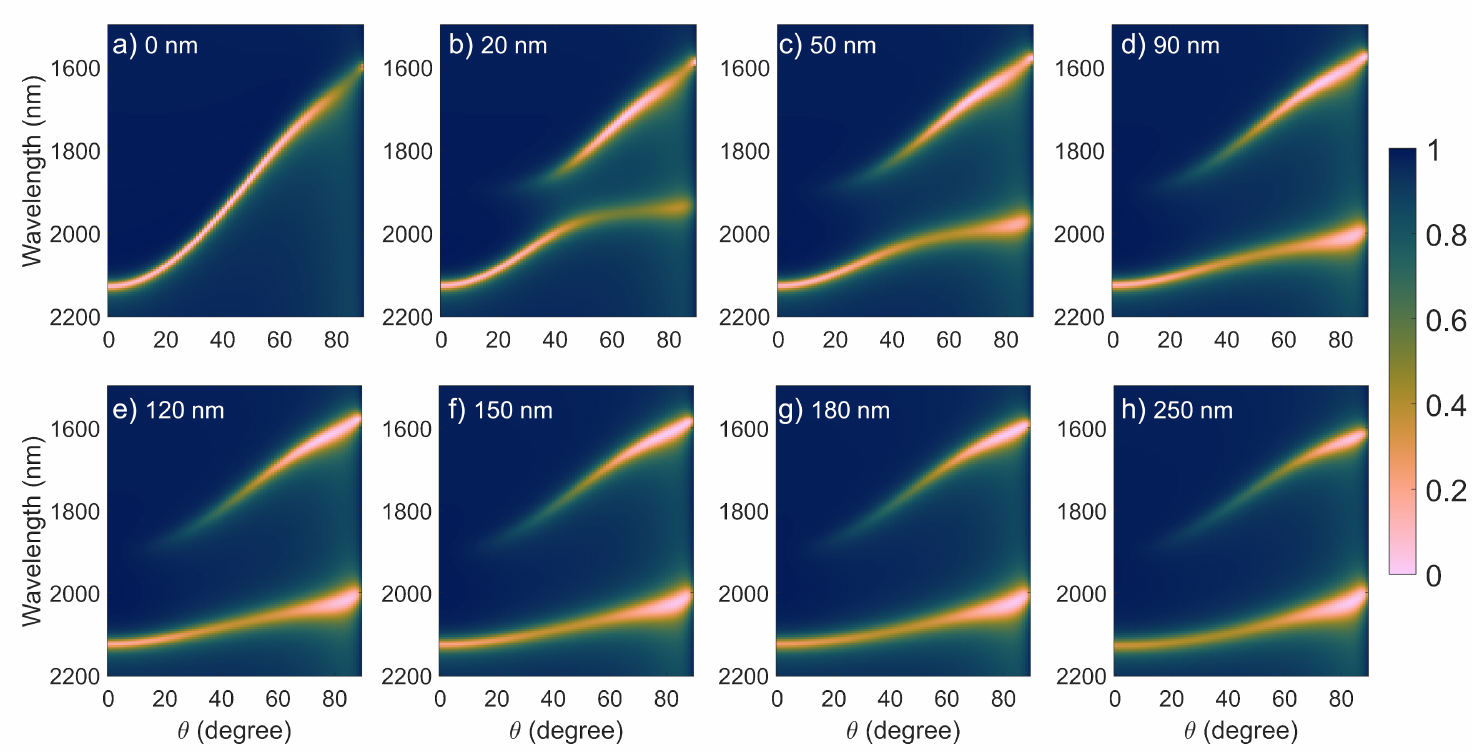}
	\caption{$R_p$ maps of the ENZ-FP cavity for different ENZ thicknesses from 0 - 250 nm, showing variation in dispersion splitting as a function of ENZ thickness. \label{Fig:3}}
\end{figure*}

\begin{figure*}[h]
	\includegraphics[width=\linewidth]{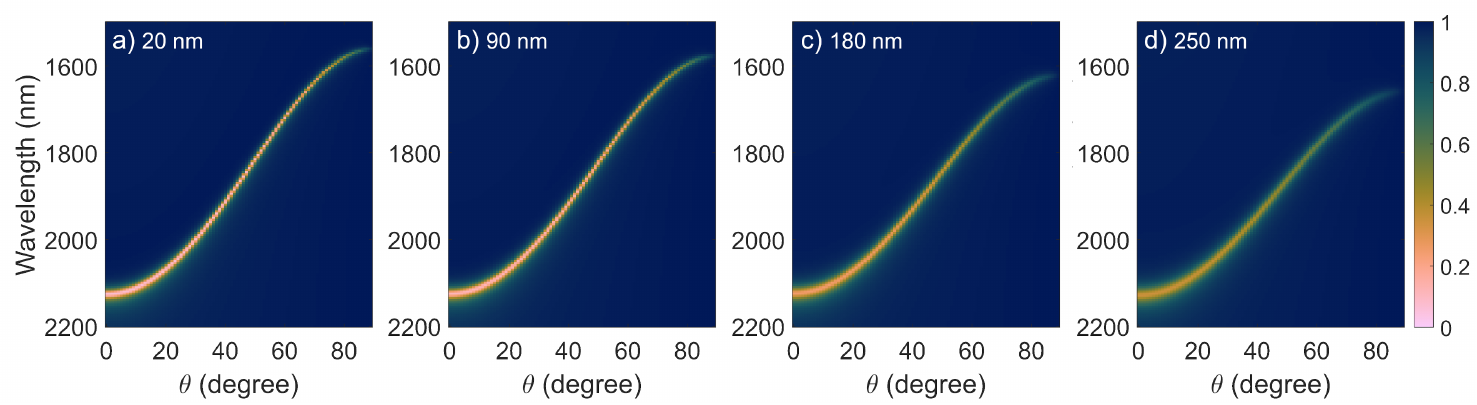}
	\caption{$R_s$ maps of the ENZ-FP cavity for four ENZ thicknesses and suitably adjusted dielectric thicknesses, showing that the effect of varying $d_{ENZ}$ on the FP dispersion is offset by suitably varying the dielectric thickness.\label{Fig:4}}
\end{figure*}

Figure \ref{Fig:3} shows the \textit{p}-polarized reflectance maps for the ENZ-FP cavity where $\lambda_{ZE}$ = 1900nm, $d_{top}$ = 20 nm, and $d_{ENZ}$ varies from 0 - 250 nm. It is evident that the spectral splitting increases initially and reaches a limiting value for $d_{ENZ} \sim$ 100 nm, beyond which the splitting remains mostly unchanged. From a practical point of view, changing $d_{ENZ}$ will also affect the spectral range of the FP resonance of the ENZ-FP cavity, which is determined by the thickness and refractive index of not only the dielectric (PMMA), but also that of the ENZ layer. In order to isolate the effect of varying $d_{ENZ}$ in the numerical calculations, any variation in the FP cavity resonance with $d_{ENZ}$ needs to be avoided. This was done by suitably adjusting the dielectric thickness such that the FP resonance of the ENZ-FP cavity remains unmodified. This ensures that only the effect of varying ENZ thickness is observed in Figure \ref{Fig:3}. To show that the FP resonance is negligibly affected and the overall spectral range of the resonance remain unchanged, Figure \ref{Fig:4} shows \textit{s}-polarized reflectance maps of the ENZ-FP cavity for four different ENZ thicknesses. The $R_s$ maps confirm that the FP dispersion is only slightly modified. The dispersions remain nearly identical even up to $d_{ENZ}$ = 250 nm and show only slight deviations at large $d_{ENZ}$, especially at very large $\theta$ values.

\section{\label{}Numerical optimization}

\begin{figure*}[h]
	\includegraphics[width=\linewidth]{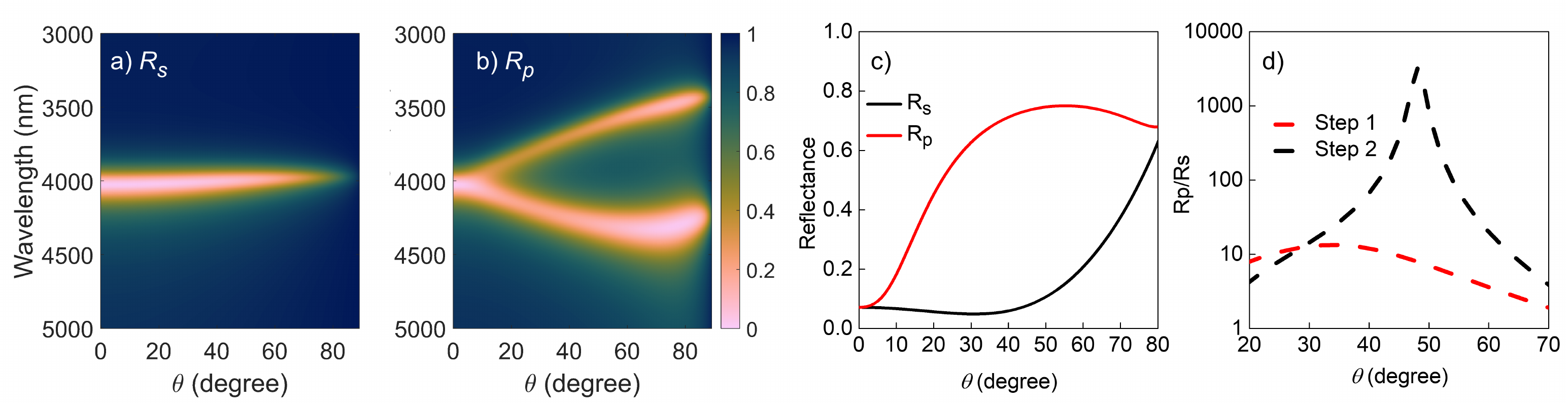}
	\caption{(a) $R_s$ and (b) $R_p$ obtained after step 1 of numerical optimization in the \textit{p}-polarizer scheme (see text). (c) Corresponding reflectance at the target operation wavelength $\lambda$ = 4000 nm. (d) Comparison of extinction ratio after step 1 and step 2 of optimization in \textit{p}-polarizer scheme. \label{Fig:5}}
\end{figure*}

The numerical optimization is carried out using a gradient-based constrained optimization routine in MATLAB \cite{fmincon}. In the \textit{s}-polarizer scheme, the quantity $R_s - R_p$ at $\lambda$ = 2100 nm is maximized simultaneously at angles in the range 15$^0$ to 70$^0$. This is done by defining the output of the objective function to be minimized as $-(R_s - R_p)$. The parameters simultaneously optimized are the layer thicknesses ($d_{top}, d$ and $d_{enz}$) and $\lambda_{ZE}$, which are reported in the main text.

For the \textit{p}-polarizer, the optimization is carried out in two steps: first, $R_p - R_s$ is maximized at $\lambda$ = 4 $\mu$m by varying the four parameters mentioned above. These values are $d_{top}$ = 173 nm, $d$ = 115 nm, $d_{enz}$ = 118 nm and $\lambda_{ZE}$ = 3.98 $\mu$m. The corresponding reflectance maps are shown in Figure \ref{Fig:5}a,b. Figure \ref{Fig:5}c also plots the reflectance contrast at $\lambda$ = 4 $\mu$m. While the reflectance contrast is maximized in this optimization step, it is evident that $R_s$ is too high for efficient wide-angle polarization. To address this limitation, a second optimization step is carried out to minimize $R_s$ where the parameters to be optimized are now only $d_{top}$ and $d$ i.e. $d_{enz}$ and $\lambda_{ZE}$ from step 1 are left unchanged. The results of step 2 are reported in the main text. To compare the results of the two steps, Figure \ref{Fig:5}d plots the extinction ratio ($R_p/R_s$) from step 1 and step 2, demonstrating a marked improvement in the figure of merit after the second optimization step.

\section{\label{}Effect of dielectric refractive index on FP dispersion}

\begin{figure*}[h]
	\includegraphics[width=\linewidth]{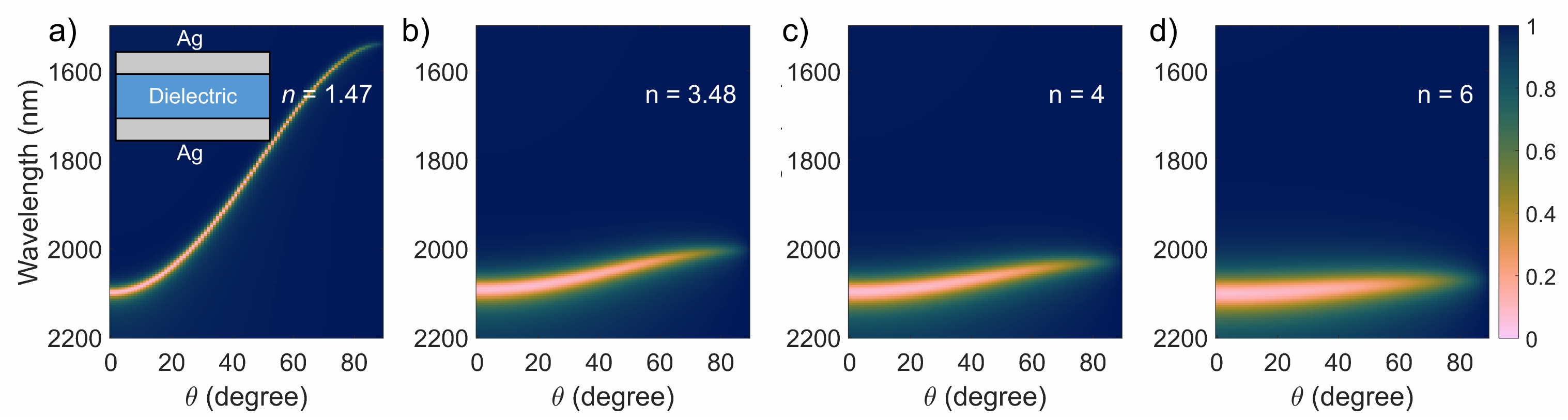}
	\caption{$R_p$ maps of the FP cavity shown in the inset for varying dielectric refractive indices as noted. Here, $d$ is chosen such that the resonant wavelength at normal incidence is fixed at 2100 nm in all cases for the sake of comparison.\label{Fig:6}}
\end{figure*}

The angle-dependence of the FP dispersion can be mitigated to an extent by using large refractive index dielectrics, which would reduce the wavelength spread of the dispersion. Typically, Si ($n \approx$ 3.4) and Ge ($n \approx$ 4) are commonly used as high-index infrared dielectrics. Figure \ref{Fig:6} plots $R_s$ for cavity dielectric refractive index values of 1.47 (PMMA), 3.48 (Si), 4 (Ge), and 6 from left to right, showing how the dispersion becomes flatter as $n$ increases. Even for $n$ = 4, the dispersion is not flat enough for practically realizing wide-angle absorption at a given wavelength, with the FP resonance wavelength varying by $\sim$ 100 nm in this case. Hence GST with $n \approx$ 6 is utilized to effectively achieve the wide-angle polarization described in the main text.
 
\section{\label{}Transfer matrix method}

\begin{figure*}[h]
	\centering
	\includegraphics[]{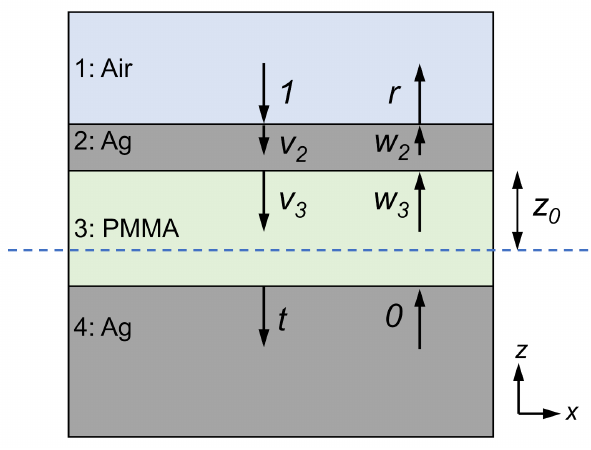}
	\caption{Schematic of representative 4-layer structure considered below. \label{Fig:7}}
\end{figure*}

In planar multi-layers, the waves in each layer may be treated as having downward (-z) and upward (+z) propagating components, as shown in a representative 4-layer system in Figure \ref{Fig:7}. The stack has $N$ media numbered 1 to $N$ where the first and last layers are semi-infinite. $v_n$ and $w_n$ denote the amplitude of downward and upward moving waves respectively. In the first and last layers, these amplitudes are 1, $r$ and $t$, 0 respectively, where $r, t$ are the overall reflection and transmission coefficients of the multi-layer. $v_1$ amd $w_1$ are not defined while $v_N = t$ and $w_N = 0$. The fields in the intermediate layers can be related as,
\begin{subequations} \label{eq:tmm}
	\begin{align}
		v_{n+1} = v_ne^{i\delta_n}t_{n,n+1} + w_{n+1}r_{n+1,n} \\
		w_n = w_{n+1}t_{n+1,n}e^{i\delta_n} + v_nr_{n,n+1}e^{2i\delta_n}
	\end{align}
\end{subequations}
where $r_{ij}$ and $t_{ij}$ are the (polarization-dependent) Fresnel reflection and transmission coefficients for light falling from layer $i$ to $j$ and $\delta_n$ is the phase term in layer $n$. This can be written in matrix form as
\begin{equation} \label{eqn:vnvn1}
	\begin{pmatrix} v_n\\ w_n \end{pmatrix} = M_n  \begin{pmatrix} v_{n+1}\\ w_{n+1} \end{pmatrix}
\end{equation}
where $n$ = 2,...,$N-1$ and \( M_n = \begin{pmatrix} e^{-i\delta_n} & 0\\ 0 & e^{i\delta_n} \end{pmatrix} \begin{pmatrix} 1 & r_{n,n+1}\\r_{n,n+1} & 1 \end{pmatrix} \frac{1}{t_{n,n+1}}\). Note that $M_1$ and $M_N$ are not defined since they are semi-infinite layers. By applying the above relation successively for each layer, one obtains the matrix $\tilde{M}$ relating the waves entering and exiting the structure (called the transfer matrix), 
\begin{equation}
	\tilde{M} = \frac{1}{t_{1,2}}\begin{pmatrix} 1 & r_{1,2}\\r_{1,2} & 1 \end{pmatrix} M_2M_3...M_{N-1}
\end{equation}
Finally, the incoming and outgoing waves are related as
\begin{equation}
	\begin{pmatrix} 1 \\ r \end{pmatrix} = \begin{pmatrix} \tilde{M_{11}} & \tilde{M_{12}} \\ \tilde{M_{21}} & \tilde{M_{22}} \end{pmatrix} \begin{pmatrix} t \\ 0 \end{pmatrix}
\end{equation}
which gives $t = 1/\tilde{M_{11}}$ and $r = \tilde{M_{21}}/\tilde{M_{11}}$ \cite{Yeh2005}.

Having calculated $(t, 0)$, which is the same as $(v_N, w_N)$, the fields in each intermediate layer $(v_n, w_n)$ can be successively calculated using Equation \eqref{eqn:vnvn1}. For example, to find the field along the dashed line in Figure \ref{Fig:7} at a distance of $z_0$ inside the PMMA layer, we have for \textit{s}-polarized light,

\begin{equation}
\label{eq:tmmfields}
	E_y(z_0) = v_3e^{ik_{z3}z_0} + w_3e^{-ik_{z3}z_0}
\end{equation}
and for \textit{p}-polarized light,
\begin{subequations} \label{eq:tmmfieldp}
	\begin{align}
		E_x(z_0) = \left[ v_3e^{ik_{z3}z_0} + w_3e^{-ik_{z3}z_0}\right] \text{cos}\, \theta_3 \\
		E_z(z_0) = \left[ -v_3e^{ik_{z3}z_0} + w_3e^{-ik_{z3}z_0}\right] \text{sin}\, \theta_3
	\end{align}
\end{subequations}
The angles $\theta_n$ in layer $n$ are calculated from Snell's law as \( \text{sin}\, \theta_n = \text{sin} \, \theta_1/\tilde{n} \) where $\theta_1$ is the angle of incidence at the first interface and $\tilde{n}$ is the complex refractive index of the considered layer. Further, \( \text{cos}\, \theta_n = \sqrt{1-\text{sin}^2 \theta_n} \) and \( k_{zn} = 2 \pi \tilde{n}\, \text{cos}\, \theta_n/\lambda\). The negative sign for $v_3$ in Equation \eqref{eq:tmmfieldp} is due to the convention where fields traveling in opposite directions have $E_z$ pointing in opposite directions \cite{griffiths_2017}.

\bibliographystyle{unsrtnat}
\bibliography{enz_FP_refs}